\documentclass{article}%
\usepackage{amsmath}
\usepackage{amsfonts}
\usepackage{amssymb}
\usepackage{graphicx}%

\begin{document}

\title{The equation of state for two-dimensional hard-sphere gases: Hard-sphere gases
as ideal gases with multi-core boundaries \thanks{A condensed version of this
paper is published in \textit{Europhys. Lett.}, \textbf{72} (6), pp. 887--893
(2005).}}
\author{Wu-Sheng Dai$^{1,2}$ \thanks{Email:daiwusheng@tju.edu.cn} and Mi Xie$^{1,2}$
\thanks{Email:xiemi@tju.edu.cn}\\{\footnotesize $^{1}$ School of Science, Tianjin University, Tianjin 300072,
P. R. China }\\{\footnotesize $^{2}$ LiuHui Center for Applied Mathematics, Nankai University
\& Tianjin University,} {\footnotesize Tianjin 300072, P. R. China} }
\date{}
\maketitle

\begin{abstract}
The equation of state for a two-dimensional hard-sphere gas is difficult to
calculate by usual methods. In this paper we develop an approach for
calculating the equation of state of hard-sphere gases, both for two- and
three-dimensional cases. By regarding a hard-sphere gas as an ideal gas
confined in a container with a multi-core (excluded sphere) boundary, we treat
the hard-sphere interaction in an interacting gas as the boundary effect on an
ideal quantum gas; this enables us to treat an interacting gas as an ideal
one. We calculate the equation of state for\ a three-dimensional hard-sphere
gas with spin $j$, and compare it with the results obtained by other methods.
By this approach the equation of state for a two-dimensional hard-sphere gas
can be calculated directly.

\end{abstract}

PACS numbers: 05.30.-d, 05.70.Ce

\vskip0.5cm

There are various methods for obtaining the equation of state of
three-dimensional hard-sphere gases, e.g., the method of pseudopotentials
\cite{HYHYL}, cluster expansions \cite{LeeYang59I,LeeYang59II}. However, it is
difficult to calculate the equation of state of two-dimensional hard-sphere
gases.\quad In this paper, for dealing with the two-dimensional hard-sphere
gases, we develop an approach for calculating the equation of state of
hard-sphere gases, both for two- and three- dimensional cases. This paper
falls into three sections. In section 1 we first introduce the basic idea of
this approach. For illustrating the approach and demonstrating its validity,
we use this approach to calculate the equation of state for three-dimensional
hard-sphere gases and compare the result with the standard result given by Lee
and Yang \cite{LeeYang59I,LeeYang59II}. In section 2 we calculate the equation
of state of two-dimensional hard-sphere gases. The paper is concluded in
section 3.

\section{Hard-sphere gases as ideal gases with multi-core boundaries: An
approach to two- and three-dimensional interacting gases}

By regarding a hard-sphere gas as an ideal gas confined in a container with a
multi-core boundary, i.e., a container filled with small excluded spheres
(Fig. 1), we calculate the grand potential for two- and three-dimensional
interacting gases using the method developed for calculating the boundary
effect on an ideal gas. The value of studying hard-sphere gases is: (1) Often
a complete knowledge of the detailed interaction potential is not necessary
for a satisfactory description of the system, because a particle that is
spread out in space sees only an averaged effect of the potential
\cite{HYHYL}. (2) Experimental progress has been made in recent years in
Bose-Einstein condensation in dilute atomic gases \cite{Anderson}; the
hard-sphere interaction model is often used as an efficient tool for studying
Bose-Einstein condensation, both for dilute atomic gases and liquid helium
\cite{Krauth}, after Lee and Yang studied pioneered the study of interacting
bosons \cite{LeeYang58}. (3) The hard-sphere gas, as a simplified model, is of
great value for investigating the more general theory of interacting gases and
can be extended to some more general cases. As a valuable idealized model, the
hard-sphere gas has been studied by various theories, such as the method of
pseudopotentials \cite{HYHYL}, cluster expansions
\cite{LeeYang59I,LeeYang59II}, and second quantization \cite{Fetter}.

The hard-sphere gas is a simplified model of the interacting gas, which
replaces the interparticle interaction by the boundary condition the
wavefunction $\psi=0$ on the boundary; such a boundary in the 3N-dimensional
configuration space is equivalent to a tree-like hypersurface
\cite{HYHYL,Huang}. In a quantum hard-sphere gas, there are two interplayed
effects: the effect of the statistics and the effect of the hard-sphere
interaction. The interplay of the effects of statistics and the interparticle
interaction often causes difficulties, e.g., it makes the quantum cluster
expansion considerably involved \cite{KU}: to find the $l$-th virial
coefficient for $l>2$, one has to solve a quantum many-body problem. There are
various approaches to hard-sphere gases. The scheme proposed by Lee and Yang,
the binary collision method, is to separate out the effect of the statistics
\cite{LeeYang59I,LeeYang59II}; it allows one first takes care of the
statistical aspect of the problem and then tackles the dynamical aspect of it
\cite{Pathria}, and the $l$-th virial coefficient can be obtained by solving
two-body problems. In the method of pseudopotentials the boundary condition is
approximately replaced by a non-Hermitian pseudopotential \cite{HYHYL,Huang},
i.e., the boundary condition is converted to a "potential" approximately.

In this paper, by replacing a hard-sphere gas by an ideal quantum gas with a
multi-core boundary (Fig. 1), and using the method developed for calculating
the boundary effects, we converted the problem of a quantum hard-sphere gas
(Fig. 1a) into a problem of an ideal quantum gas confined in a container
filled with small excluded spheres (cores) randomly distributed (Fig. 1b). In
this approach, roughly speaking, the exchange effect and the effect of
classical interaction (in this case, the hard-sphere interaction) are treated
separately: the exchange effect on the imperfect gas is embodied in the
exchange effect on the ideal gas; the effect of hard-sphere interaction is
embodied in the boundary effect. The number of the cores is chosen to equal
the total number of particles; the radius of the cores is chosen to equal the
diameter of the hard sphere $a$ (the scattering length of the hard sphere)
since the area of a core felt by a gas molecule during the scattering process
is $4\pi a^{2}$.\ \ \ \ \begin{figure}[h]
\begin{center}
{\includegraphics{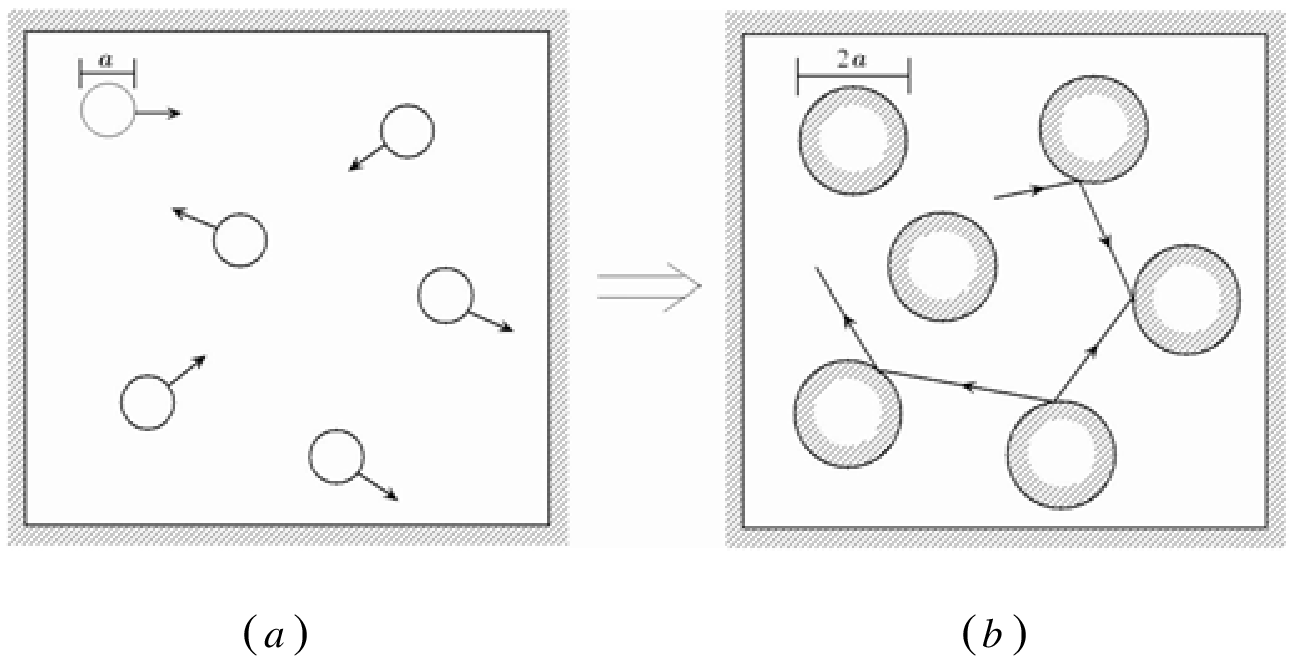}}
\end{center}
\par
{\small Fig. 1. Regarding a hard-sphere gas as an ideal quantum gas in a
multi-core container. (a) A hard-sphere gas in which the diameter of the gas
molecule is $a$. (b) An ideal quantum gas confined in a container with a
multi-core boundary. The diameter of the core is $2a$.}\end{figure}

Then, the task that to calculate an interacting gas is converted to a task
that to calculate the boundary effect on an ideal quantum gas. In preceding
papers we developed some methods for calculating boundary effects
\cite{Ours1,Ours2}: one \cite{Ours1}, which is based on the mathematical work
given by Kac \cite{Kac}, is for two-dimensional cases; the other \cite{Ours2}
is for three-dimensional cases. However, to apply these methods to the present
problem, in which the boundary is more complex, we need to extend them to more
general cases \cite{Ours3}: the grand potentials for ideal quantum gases in
three- and two- dimensional confined space can be expressed as
\begin{align}
\ln\Xi &  =g\frac{1}{2^{0}}\frac{\mu_{3}^{o}-\left(  -1\right)  ^{d-3}\mu
_{3}^{i}}{\lambda^{3}}h_{5/2}\left(  z\right)  -g\frac{1}{2^{1}}\frac{\mu
_{2}^{o}-\left(  -1\right)  ^{d-2}\mu_{2}^{i}}{\lambda^{2}}h_{2}\left(
z\right)  +g\frac{1}{2^{2}}\frac{\mu_{1}^{o}-\left(  -1\right)  ^{d-1}\mu
_{1}^{i}}{\lambda^{1}}h_{3/2}\left(  z\right)  ,\text{ (3D)}\\
\ln\Xi &  =g\frac{1}{2^{0}}\frac{\mu_{2}^{o}-\left(  -1\right)  ^{d-2}\mu
_{2}^{i}}{\lambda^{2}}h_{2}\left(  z\right)  -g\frac{1}{2^{1}}\frac{\mu
_{1}^{o}-\left(  -1\right)  ^{d-1}\mu_{1}^{i}}{\lambda^{1}}h_{3/2}\left(
z\right)  ,\text{ \ \ \ \ \ \ \ \ \ \ \ \ \ \ \ \ \ \ \ \ \ \ \ (2D)}%
\end{align}
where $d$ denotes the dimension; $\lambda=h/\sqrt{2\pi mkT}$ is the thermal
wavelength, $g$ the number of internal degrees of freedom and for spins
$g=2j+1$, and the function $h_{\sigma}(z)=\frac{1}{\Gamma(\sigma)}\int
_{0}^{\infty}\frac{x^{\sigma-1}}{z^{-1}e^{x}\mp1}dx$ equals the\ Bose-Einstein
integral $g_{\sigma}(z)$ or the Fermi-Dirac integral $f_{\sigma}(z)$ in boson
or fermion case respectively. $\mu_{\alpha}^{o}$ and $\mu_{\alpha}^{i}$ are
the \textit{valuations} of the containers \cite{KR} ($\mu_{\alpha}^{o}$ for
outer boundaries and $\mu_{\alpha}^{i}$ for inner boundaries): in three
dimensions for a sphere $\mu_{3}$ is the volume, $\mu_{2}$ half the surface
area, and $\mu_{1}$ twice the diameter; in two dimensions, for a disk $\mu
_{2}$ is the area, and $\mu_{1}$ half the perimeter. However, this result is
only valid for spinless distinguishable particles. To take the effects of spin
and indistinguishability into account, we need to improve this result one step
further. The hard-sphere interaction is embodied in the boundary terms; the
cross-section for collisions between gas molecules is regarded as the surface
area of the boundary cores. For distinguishable hard-sphere particles, the
cross-section is $4\pi a^{2}$; however, for indistinguishable ones, the
cross-section is $8\pi a^{2}$, i.e., the contribution from the
indistinguishable hard-sphere interaction is twice the contribution from
distinguishable ones. Moreover, for the case of particles with spin $j$, if we
include only the $s$-wave contribution, there are weights $\left(  j+1\right)
/\left(  2j+1\right)  $ for Bosons and $j/\left(  2j+1\right)  $ for Fermions
respectively for the unpolarized scattering of identical particles
\cite{Bernstein}. Thus the grand potential of an ideal quantum gas confined in
the container illustrated in Fig. 1b can be written in the following form:%

\begin{equation}
\ln\Xi=g\frac{V}{\lambda^{3}}h_{5/2}\left(  z\right)  +g2\frac{\omega}%
{2j+1}\left[  -\frac{1}{2}\frac{\frac{1}{2}N4\pi a^{2}}{\lambda^{2}}%
h_{2}\left(  z\right)  +\frac{1}{4}\frac{-2N2a}{\lambda}h_{3/2}\left(
z\right)  \right]  ,\label{1}%
\end{equation}
where $\omega=j+1$ for bosons and $\omega=j$ for fermions, $a$ is the diameter
of the hard sphere and $N$ is the total number of particles in the system.
Note that the boundary, of course, includes all cores. Strictly speaking, see
Fig. 1, the volume of the system is the volume of the cubical container $V$
minus the total volume of all cores; however, in view of the fact that for a
dilute system the former is much larger than the latter and the contribution
from the volume of the cores is only of the order $\left(  a/\lambda\right)
^{3}$, we approximately use $V$ as the volume of the system. In the second
term the factor $\left(  1/2\right)  N4\pi a^{2}$ is half the total surface
area of all cores --- the second valuation, which is much larger than the
magnitude of $V^{2/3}$ in a macroscopic system; in the third term the factor
$2\times2a$ is twice the diameter of a core so $2N2a$ is the third valuation
of the boundary (notice that $2N2a\gg V^{1/3}$). The factors $2$ and
$\omega/\left(  2j+1\right)  $ reflect the effects of indistinguishability and
spin. From Eq. (\ref{1}), the relation $N=z\partial\ln\Xi/\partial z$ gives a
differential equation in $N$. Such an equation for $N$ is solvable (see
Appendix); however, its solution is in a complicated form. In the following,
we present a series solution, which is easy to be compared with the results
given in the literature, and, of course, it can be checked directly that such
a series solution is consistent with the solution obtained by directly solving
the differential equation for $N$. For seeking a series solution for $N$, we
express $N$ as a combination of the Bose-Einstein or the Fermi-Dirac
integrals, $h_{\sigma}(z)$,%
\begin{align}
N &  =g\frac{V}{\lambda^{3}}\Bigg\{\left[  \sum\limits_{\mu}A_{\mu}h_{\mu
}+\sum\limits_{\nu\leq\sigma}B_{\nu\sigma}h_{\nu}h_{\sigma}+\sum
\limits_{\xi\leq\eta\leq\lambda}C_{\xi\eta\lambda}h_{\xi}h_{\eta}h_{\lambda
}+\cdots\right]  \nonumber\\
&  +\frac{a}{\lambda}\left[  \sum\limits_{\mu}A_{\mu}^{\prime}h_{\mu}%
+\sum\limits_{\nu\leq\sigma}B_{\nu\sigma}^{\prime}h_{\nu}h_{\sigma}%
+\sum\limits_{\xi\leq\eta\leq\lambda}C_{\xi\eta\lambda}^{\prime}h_{\xi}%
h_{\eta}h_{\lambda}+\cdots\right]  \nonumber\\
&  +\left(  \frac{a}{\lambda}\right)  ^{2}\left[  \sum\limits_{\mu}A_{\mu
}^{\prime\prime}h_{\mu}+\sum\limits_{\nu\leq\sigma}B_{\nu\sigma}^{\prime
\prime}h_{\nu}h_{\sigma}+\sum\limits_{\xi\leq\eta\leq\lambda}C_{\xi\eta
\lambda}^{\prime\prime}h_{\xi}h_{\eta}h_{\lambda}+\cdots\right]
+\cdots\Bigg\},\label{2}%
\end{align}
where the sum is taken by all possible $h_{\sigma}(z)$ and the subscripts of
the coefficients are taken in ascending order for avoiding repetition, and
substituting Eq. (\ref{2}) into Eq. (\ref{1}), accurate to $\left(
a/\lambda\right)  ^{2}$, we obtain%
\begin{align}
\ln\Xi &  =g\frac{V}{\lambda^{3}}\left\{  h_{5/2}-\frac{a}{\lambda}%
2g\frac{\omega}{2j+1}\left[  \sum\limits_{\mu}A_{\mu}h_{3/2}h_{\mu}%
+\sum\limits_{\nu\leq\sigma}B_{\nu\sigma}h_{3/2}h_{\nu}h_{\sigma}%
+\sum\limits_{\xi\leq\eta\leq\lambda}C_{\xi\eta\lambda}h_{3/2}h_{\xi}h_{\eta
}h_{\lambda}\right]  \right.  \nonumber\\
&  -\left(  \frac{a}{\lambda}\right)  ^{2}2g\frac{\omega}{2j+1}\left[
\sum\limits_{\mu}\pi A_{\mu}h_{2}h_{\mu}+\sum\limits_{\nu\leq\sigma}\pi
B_{\nu\sigma}h_{2}h_{\nu}h_{\sigma}+\sum\limits_{\xi\leq\eta\leq\lambda}\pi
C_{\xi\eta\lambda}h_{2}h_{\xi}h_{\eta}h_{\lambda}\right.  \nonumber\\
&  \left.  -\left.  \sum\limits_{\mu}A_{\mu}^{\prime}h_{3/2}h_{\mu}%
-\sum\limits_{\nu\leq\sigma}B_{\nu\sigma}^{\prime}h_{3/2}h_{\nu}h_{\sigma
}-\sum\limits_{\xi\leq\eta\leq\lambda}C_{\xi\eta\lambda}^{\prime}h_{3/2}%
h_{\xi}h_{\eta}h_{\lambda}\right]  \right\}  .
\end{align}
Deriving $N$ by using the relations $N=z\partial\ln\Xi/\partial z$ and
$z\partial h_{\sigma}/\partial z=h_{\sigma-1}$, after ignoring higher-order
terms, we have%
\begin{align*}
N &  =g\frac{V}{\lambda^{3}}\Bigg\{h_{3/2}\\
&  -\frac{a}{\lambda}2g\frac{\omega}{2j+1}\left[  \sum\limits_{\mu}A_{\mu
}h_{1/2}h_{\mu}+\sum\limits_{\mu}A_{\mu}h_{3/2}h_{\mu-1}+\sum\limits_{\nu
\leq\sigma}B_{\nu\sigma}h_{1/2}h_{\nu}h_{\sigma}+\sum\limits_{\nu\leq\sigma
}B_{\nu\sigma}h_{3/2}h_{\nu-1}h_{\sigma}+\sum\limits_{\nu\leq\sigma}%
B_{\nu\sigma}h_{3/2}h_{\nu}h_{\sigma-1}\right]  \\
&  -\left(  \frac{a}{\lambda}\right)  ^{2}2g\frac{\omega}{2j+1}\left[
\sum\limits_{\mu}\pi A_{\mu}h_{1}h_{\mu}+\sum\limits_{\mu}\pi A_{\mu}%
h_{2}h_{\mu-1}+\sum\limits_{\nu\leq\sigma}\pi B_{\nu\sigma}h_{1}h_{\nu
}h_{\sigma}+\sum\limits_{\nu\leq\sigma}\pi B_{\nu\sigma}h_{2}h_{\nu
-1}h_{\sigma}+\sum\limits_{\nu\leq\sigma}\pi B_{\nu\sigma}h_{2}h_{\nu
}h_{\sigma-1}\right.  \\
&  \left.  +\sum\limits_{\mu}A_{\mu}^{\prime}h_{1/2}h_{\mu}+\sum\limits_{\mu
}A_{\mu}^{\prime}h_{3/2}h_{\mu-1}+\sum\limits_{\nu\leq\sigma}B_{\nu\sigma
}^{\prime}h_{1/2}h_{\nu}h_{\sigma}+\sum\limits_{\nu\leq\sigma}B_{\nu\sigma
}^{\prime}h_{3/2}h_{\nu-1}h_{\sigma}+\sum\limits_{\nu\leq\sigma}B_{\nu\sigma
}^{\prime}h_{3/2}h_{\nu}h_{\sigma-1}\right]  \Bigg\}
\end{align*}
and comparing the expression of $N$ with Eq. (\ref{2}), we obtain the
coefficients: $A_{3/2}=1$, $B_{1/2,3/2}^{\prime}=-4g\omega/\left(
2j+1\right)  $, $B_{1,3/2}^{\prime\prime}=B_{1/2,2}^{\prime\prime}=-2\pi
g\omega/\left(  2j+1\right)  $, $C_{1/2,1/2,3/2}^{\prime\prime}%
=2C_{-1/2,3/2,3/2}^{\prime\prime}=16g^{2}\left[  \omega/\left(  2j+1\right)
\right]  ^{2}$, and the other coefficients equal $0$. Then, we achieve the
equation of state, Eq. (\ref{5})%
\begin{equation}
\lambda^{3}\frac{P}{kT}=\left(  2j+1\right)  h_{5/2}-2\omega\left(
2j+1\right)  \frac{a}{\lambda}h_{3/2}^{2}+\left(  \frac{a}{\lambda}\right)
^{2}\left[  8\omega^{2}\left(  2j+1\right)  h_{1/2}h_{3/2}^{2}-\omega\left(
2j+1\right)  2\pi h_{3/2}h_{2}\right]  .\label{5}%
\end{equation}
Here we have substituted the relation $g=2j+1$ and replaced the Bose-Einstein
integral $g_{\sigma}\left(  z\right)  $\ by $h_{\sigma}\left(  z\right)
$,\ which includes the Fermi case. This is the result for an ideal gas
confined in the container, Fig. 1b, and this is also the result what we want
to obtain for a hard-sphere gas, Fig. 1a.

Alternatively, we can also obtain this result in a more rigorous way. For
seeking the series solution for $N$, we can also express $N$ as
\begin{equation}
N=g\frac{V}{\lambda^{3}}\left[  \sum\limits_{\mu=1}^{\infty}A_{\mu}z^{\mu
}+\frac{a}{\lambda}\sum\limits_{\nu=1}^{\infty}B_{\nu}z^{\nu}+\left(  \frac
{a}{\lambda}\right)  ^{2}\sum\limits_{\sigma=1}^{\infty}C_{\sigma}z^{\sigma
}+\cdots\right]  .\label{i01}%
\end{equation}
In the following we take the Bose-Einstein case as an example, and the result
for the Fermi-Dirac case can be achieved by a very similar procedure.
Substituting Eq. (\ref{i01}) into Eq. (\ref{1}) and expanding the
Bose-Einstein integral as a series by use of $g_{\sigma}=\sum_{l=1}^{\infty
}z^{l}/l^{\sigma}$, we have%
\begin{align}
\ln\Xi &  =g\frac{V}{\lambda^{3}}\left\{  \sum_{l=1}^{\infty}\frac{z^{l}%
}{l^{5/2}}-\frac{a}{\lambda}2g\frac{j+1}{2j+1}\left[  \sum_{\mu,l=1}^{\infty
}\frac{A_{\mu}}{l^{3/2}}z^{\mu+l}+\frac{a}{\lambda}\sum_{\mu,l=1}^{\infty
}\frac{B_{\mu}}{l^{3/2}}z^{\mu+l}+\left(  \frac{a}{\lambda}\right)  ^{2}%
\sum_{\mu,l=1}^{\infty}\frac{C_{\mu}}{l^{3/2}}z^{\mu+l}\right]  \right.
\nonumber\\
&  \left.  -\left(  \frac{a}{\lambda}\right)  ^{2}\pi2g\frac{j+1}{2j+1}\left[
\sum_{\mu,l=1}^{\infty}\frac{A_{\mu}}{l^{2}}z^{\mu+l}+\frac{a}{\lambda}%
\sum_{\mu,l=1}^{\infty}\frac{B_{\mu}}{l^{2}}z^{\mu+l}+\left(  \frac{a}%
{\lambda}\right)  ^{2}\sum_{\mu,l=1}^{\infty}\frac{C_{\mu}}{l^{2}}z^{\mu
+l}\right]  \right\}  .\label{seq}%
\end{align}
The relation $N=z\partial\ln\Xi/\partial z$, after ignoring higher-order
terms, gives%
\begin{align*}
N  &  =g\frac{V}{\lambda^{3}}\left\{  \sum_{l=1}^{\infty}l\frac{z^{l}}%
{l^{5/2}}-2g\frac{j+1}{2j+1}\frac{a}{\lambda}\sum_{\mu,l=1}^{\infty}\left(
\mu+l\right)  \frac{A_{\mu}}{l^{3/2}}z^{\mu+l}-2g\frac{j+1}{2j+1}\left(
\frac{a}{\lambda}\right)  ^{2}\sum_{\mu,l=1}^{\infty}\left(  \mu+l\right)
\frac{B_{\mu}}{l^{3/2}}z^{\mu+l}\right. \\
&  \left.  -2\pi g\frac{j+1}{2j+1}\left(  \frac{a}{\lambda}\right)  ^{2}%
\sum_{\mu,l=1}^{\infty}\left(  \mu+l\right)  \frac{A_{\mu}}{l^{2}}z^{\mu
+l}\right\} \\
&  =g\frac{V}{\lambda^{3}}\left\{  \sum_{\mu=1}^{\infty}\frac{z^{\mu}}%
{\mu^{3/2}}-2g\frac{j+1}{2j+1}\left[  \frac{a}{\lambda}\sum_{\mu=2}^{\infty
}\sum\limits_{k=1}^{\mu-1}\frac{\mu A_{k}}{\left(  \mu-k\right)  ^{3/2}}%
z^{\mu}-\left(  \frac{a}{\lambda}\right)  ^{2}\sum_{\mu=2}^{\infty}%
\sum\limits_{k=1}^{\mu-1}\left(  \frac{-\mu\pi A_{k}}{\left(  \mu-k\right)
^{2}}-\frac{\mu B_{k}}{\left(  \mu-k\right)  ^{3/2}}\right)  z^{\mu}\right]
\right\}  .
\end{align*}
Comparing with Eq. (\ref{i01}), we obtain%
\begin{align*}
A_{\mu}  &  =\frac{1}{\mu^{3/2}},\text{ \ }B_{\mu}=-2g\frac{j+1}{2j+1}%
\sum\limits_{k=1}^{\mu-1}\frac{\mu}{\left[  \left(  \mu-k\right)  k\right]
^{3/2}},\\
C_{\mu}  &  =2g\frac{j+1}{2j+1}\left[  \sum\limits_{k=1}^{\mu-1}\frac{-\pi\mu
}{\left(  \mu-k\right)  ^{2}k^{3/2}}+2g\frac{j+1}{2j+1}\sum\limits_{k=2}%
^{\mu-1}\sum\limits_{l=1}^{k-1}\frac{\mu}{\left(  \mu-k\right)  ^{3/2}}%
\frac{k}{\left[  \left(  k-l\right)  l\right]  ^{3/2}}\right]  ,
\end{align*}
where the summation $\sum\nolimits_{a}^{b}$ should be regarded as $0$ when
$b<a$. Substituting these parameters into Eq. (\ref{seq}) gives%
\begin{align*}
\ln\Xi &  =g\frac{V}{\lambda^{3}}\left\{  \sum_{l=1}^{\infty}\frac{z^{l}%
}{l^{5/2}}+2g\frac{j+1}{2j+1}\left[  -\frac{a}{\lambda}\sum_{\mu,l=1}^{\infty
}\frac{1}{l^{3/2}}\frac{1}{\mu^{3/2}}z^{\mu+l}\right.  \right. \\
&  \left.  \left.  +\left(  \frac{a}{\lambda}\right)  ^{2}\left(  2g\frac
{j+1}{2j+1}2\sum_{\mu,l,k=1}^{\infty}\frac{1}{l^{3/2}}\frac{1}{\mu^{1/2}}%
\frac{1}{k^{3/2}}z^{\mu+k+l}-\pi\sum_{\mu,l=1}^{\infty}\frac{1}{l^{2}}\frac
{1}{\mu^{3/2}}z^{\mu+l}\right)  \right]  \right\}  .
\end{align*}
Performing the summations and substituting the relation $g=2j+1$, we again
achieve the equation of state, Eq. (\ref{5}).

Our result Eq. (\ref{5}) compares well with the result based on the binary
collision method given by Lee and Yang \cite{LeeYang59II},%
\begin{equation}
\lambda^{3}\frac{P}{kT}=\left(  2j+1\right)  h_{5/2}-2\omega\left(
2j+1\right)  \frac{a}{\lambda}h_{3/2}^{2}+\left(  \frac{a}{\lambda}\right)
^{2}\left[  8\omega^{2}\left(  2j+1\right)  h_{1/2}h_{3/2}^{2}+\omega\left(
2j+1\right)  8F\left(  \pm z\right)  \right]  ,\label{6}%
\end{equation}
where $F\left(  z\right)  =\sum_{r,s,t=1}^{\infty}\left(  rst\right)
^{-1/2}\left(  r+s\right)  ^{-1}\left(  r+t\right)  ^{-1}z^{r+s+t}$, "$+$" for
bosons and "$-$" for fermions. The first-order contributions, $a/\lambda$, are
the same; the difference only appears in the second term of the second-order,
$\left(  a/\lambda\right)  ^{2}$, contribution. To show this difference
clearly, as an example, we compare them in the low-temperature and
high-density Fermi case, which is the most interesting case --- the quantum
case. The asymptotic forms for large $z$ are $8F\left(  -z\right)
\sim-2.10\left(  \ln z\right)  ^{7/2}$ and $-2\pi h_{3/2}h_{2}\sim-2.36\left(
\ln z\right)  ^{7/2}$. For Bose cases, at the high temperatures and low
temperatures, our result is consistent with Lee and Yang's. At low
temperatures and high densities, it is physically meaningless to compare the
results Eqs. (\ref{5}) and (\ref{6}) since they are both invalid due to the
existence of the critical point; in fact in this case the perturbation theory
in $a$ makes no sense \cite{LeeYang58,BBHLV}.\ Notice that the methods given
in Ref. \cite{LeeYang59II} and in this paper are both approximate methods
giving fugacity series of grand potentials.

\section{The equation of state for two-dimensional hard sphere gases}

The two-dimensional hard-sphere quantum gases which are difficult to calculate
by usual methods can also be calculated by this approach. Like the
three-dimensional case, the grand potential for the two-dimensional cases can
be expressed as%
\begin{align}
\ln\Xi &  =g\frac{1}{\lambda^{2}}\left(  S-2\frac{\omega}{2j+1}N\pi
a^{2}\right)  h_{2}-g\frac{1}{2}\frac{1}{\lambda}\left(  2\frac{\omega}%
{2j+1}N\frac{1}{2}2\pi a\right)  h_{3/2}\nonumber\\
&  =g\frac{S}{\lambda^{2}}h_{2}-g2\frac{\omega}{2j+1}\frac{a}{\lambda}%
\frac{\pi}{2}Nh_{3/2}-g2\frac{\omega}{2j+1}\left(  \frac{a}{\lambda}\right)
^{2}\pi Nh_{2}.\label{7}%
\end{align}
It should be emphasized that, in the two-dimensional case we take the area of
the cores into account since it provides a second-order, $\left(
a/\lambda\right)  ^{2}$, contribution (Recall that in the three-dimensional
case we ignore the contribution coming from the volume of the cores since the
volume contribution is of the order $\left(  a/\lambda\right)  ^{3}$).

By analyzing the grand potential Eq. (\ref{7}), we express the particle number
$N$ as%
\begin{align}
N  &  =g\frac{S}{\lambda^{2}}\left\{  \left[  \sum\limits_{\mu}A_{\mu}h_{\mu
}+\sum\limits_{\nu\leq\sigma}B_{\nu\sigma}h_{\nu}h_{\sigma}+\sum
\limits_{\xi\leq\eta\leq\lambda}C_{\xi\eta\lambda}h_{\xi}h_{\eta}h_{\lambda
}+\cdots\right]  \right. \nonumber\\
&  +\frac{a}{\lambda}\left[  \sum\limits_{\mu}A_{\mu}^{\prime}h_{\mu}%
+\sum\limits_{\nu\leq\sigma}B_{\nu\sigma}^{\prime}h_{\nu}h_{\sigma}%
+\sum\limits_{\xi\leq\eta\leq\lambda}C_{\xi\eta\lambda}^{\prime}h_{\xi}%
h_{\eta}h_{\lambda}+\cdots\right] \nonumber\\
&  \left.  +\left(  \frac{a}{\lambda}\right)  ^{2}\left[  \sum\limits_{\mu
}A_{\mu}^{\prime\prime}h_{\mu}+\sum\limits_{\nu\leq\sigma}B_{\nu\sigma
}^{\prime\prime}h_{\nu}h_{\sigma}+\sum\limits_{\xi\leq\eta\leq\lambda}%
C_{\xi\eta\lambda}^{\prime\prime}h_{\xi}h_{\eta}h_{\lambda}+\cdots\right]
+\cdots\right\}  .\label{i02}%
\end{align}
Substituting $N$ into Eq. (\ref{7}), we have%
\begin{align}
\ln\Xi &  =g\frac{S}{\lambda^{2}}h_{2}-g\frac{S}{\lambda^{2}}\pi g\frac
{\omega}{2j+1}\frac{a}{\lambda}\left[  \sum\limits_{\mu}A_{\mu}h_{3/2}h_{\mu
}+\sum\limits_{\nu\leq\sigma}B_{\nu\sigma}h_{3/2}h_{\nu}h_{\sigma}%
+\sum\limits_{\xi\leq\eta\leq\lambda}C_{\xi\eta\lambda}h_{3/2}h_{\xi}h_{\eta
}h_{\lambda}\right] \nonumber\\
&  -g\frac{S}{\lambda^{2}}\pi g\frac{\omega}{2j+1}\left(  \frac{a}{\lambda
}\right)  ^{2}\left[  \sum\limits_{\mu}A_{\mu}^{\prime}h_{3/2}h_{\mu}%
+\sum\limits_{\nu\leq\sigma}B_{\nu\sigma}^{\prime}h_{3/2}h_{\nu}h_{\sigma
}+\sum\limits_{\xi\leq\eta\leq\lambda}C_{\xi\eta\lambda}^{\prime}h_{3/2}%
h_{\xi}h_{\eta}h_{\lambda}\right] \nonumber\\
&  -g\frac{S}{\lambda^{2}}2\pi g\frac{\omega}{2j+1}\left(  \frac{a}{\lambda
}\right)  ^{2}\left[  \sum\limits_{\mu}A_{\mu}h_{2}h_{\mu}+\sum\limits_{\nu
\leq\sigma}B_{\nu\sigma}h_{2}h_{\nu}h_{\sigma}+\sum\limits_{\xi\leq\eta
\leq\lambda}C_{\xi\eta\lambda}h_{2}h_{\xi}h_{\eta}h_{\lambda}\right]
.\label{i03}%
\end{align}
By $N=$ $z\partial\ln\Xi/\partial z$, after Ignoring higher-order terms, we
have%
\begin{align*}
N  &  =g\frac{S}{\lambda^{2}}h_{1}\\
&  -g\frac{S}{\lambda^{2}}\pi g\frac{\omega}{2j+1}\frac{a}{\lambda}\left[
\sum\limits_{\mu}A_{\mu}h_{1/2}h_{\mu}+\sum\limits_{\mu}A_{\mu}h_{3/2}%
h_{\mu-1}+\sum\limits_{\nu\leq\sigma}B_{\nu\sigma}h_{1/2}h_{\nu}h_{\sigma
}+\sum\limits_{\nu\leq\sigma}B_{\nu\sigma}h_{3/2}h_{\nu-1}h_{\sigma}%
+\sum\limits_{\nu\leq\sigma}B_{\nu\sigma}h_{3/2}h_{\nu}h_{\sigma-1}\right] \\
&  -g\frac{S}{\lambda^{2}}\pi g\frac{\omega}{2j+1}\left(  \frac{a}{\lambda
}\right)  ^{2}\left[  \sum\limits_{\mu}A_{\mu}^{\prime}h_{1/2}h_{\mu}%
+\sum\limits_{\mu}A_{\mu}^{\prime}h_{3/2}h_{\mu-1}+\sum\limits_{\nu\leq\sigma
}B_{\nu\sigma}^{\prime}h_{1/2}h_{\nu}h_{\sigma}+\sum\limits_{\nu\leq\sigma
}B_{\nu\sigma}^{\prime}h_{3/2}h_{\nu-1}h_{\sigma}+\sum\limits_{\nu\leq\sigma
}B_{\nu\sigma}^{\prime}h_{3/2}h_{\nu}h_{\sigma-1}\right] \\
&  -g\frac{S}{\lambda^{2}}2\pi g\frac{\omega}{2j+1}\left(  \frac{a}{\lambda
}\right)  ^{2}\left[  \sum\limits_{\mu}A_{\mu}h_{1}h_{\mu}+\sum\limits_{\mu
}A_{\mu}h_{2}h_{\mu-1}+\sum\limits_{\nu\leq\sigma}B_{\nu\sigma}h_{1}h_{\nu
}h_{\sigma}+\sum\limits_{\nu\leq\sigma}B_{\nu\sigma}h_{2}h_{\nu-1}h_{\sigma
}+\sum\limits_{\nu\leq\sigma}B_{\nu\sigma}h_{2}h_{\nu}h_{\sigma-1}\right]  .
\end{align*}
Comparing the coefficients with the corresponding coefficients in (\ref{i02}),
we have%
\begin{align*}
A_{1}  &  =1,A_{\mu}=0,B_{\nu\sigma}=0,C_{\xi\eta\lambda}=0\\
A_{\mu}^{\prime}  &  =0,B_{1/2,1}^{\prime}=-\pi g\frac{\omega}{2j+1}%
,B_{0,3/2}^{\prime}=-\pi g\frac{\omega}{2j+1},B_{\nu\sigma}^{\prime}%
=0,C_{\xi\eta\lambda}^{\prime}=0\\
A_{\mu}^{\prime\prime}  &  =0,B_{1,1}^{\prime\prime}=-2\pi g\frac{\omega
}{2j+1},B_{0,2}^{\prime\prime}=-2\pi g\frac{\omega}{2j+1},B_{\nu\sigma
}^{\prime\prime}=0,\\
C_{1/2,1/2,1}^{\prime\prime}  &  =\left(  \pi g\frac{\omega}{2j+1}\right)
^{2},C_{0,1/2,3/2}^{\prime\prime}=3\left(  \pi g\frac{\omega}{2j+1}\right)
^{2},C_{-1/2,1,3/2}^{\prime\prime}=\left(  \pi g\frac{\omega}{2j+1}\right)
^{2},\\
C_{-1,3/2,3/2}^{\prime\prime}  &  =\left(  \pi g\frac{\omega}{2j+1}\right)
^{2},C_{\xi\eta\lambda}^{\prime\prime}=0.
\end{align*}
Then we obtain the equation of state for two-dimensional gases ($g=2j+1)$:%
\begin{equation}
\lambda^{2}\frac{P}{kT}=\left(  2j+1\right)  h_{2}+\omega\left(  2j+1\right)
\left[  -\frac{a}{\lambda}\pi h_{1}h_{3/2}+\left(  \frac{a}{\lambda}\right)
^{2}\left(  \omega\pi^{2}h_{0}h_{3/2}^{2}+\omega\pi^{2}h_{1/2}h_{1}%
h_{3/2}-2\pi h_{1}h_{2}\right)  \right]  .\label{i06}%
\end{equation}

Alternatively we can also obtain the equation of state in a more rigorous way.
For seeking a series solution for $N$, we expand $N$ as%
\begin{equation}
N=g\frac{S}{\lambda^{2}}\left[  \sum\limits_{\mu}A_{\mu}z^{\mu}+\frac
{a}{\lambda}\sum\limits_{\mu}B_{\mu}z^{\mu}+\left(  \frac{a}{\lambda}\right)
^{2}\sum\limits_{\mu}C_{\mu}z^{\mu}+\cdots\right]  .\label{8}%
\end{equation}
Substituting $N$ into Eq. (\ref{7}), taking the Bose case as an example, we
have ($\omega=j+1$)%
\begin{align}
\ln\Xi &  =g\frac{S}{\lambda^{2}}\sum_{l=1}^{\infty}\frac{z^{l}}{l^{2}}%
-g\frac{S}{\lambda^{2}}\frac{a}{\lambda}\left(  \pi g\frac{j+1}{2j+1}\right)
\sum_{\mu,l=1}^{\infty}A_{\mu}\frac{z^{\mu+l}}{l^{3/2}}\nonumber\\
&  -g\frac{S}{\lambda^{2}}\left(  \frac{a}{\lambda}\right)  ^{2}\left(  \pi
g\frac{j+1}{2j+1}\right)  \sum_{\mu,l=1}^{\infty}\left[  B_{\mu}\frac
{z^{\mu+l}}{l^{3/2}}+2A_{\mu}\frac{z^{\mu+l}}{l^{2}}\right] \nonumber\\
&  =g\frac{S}{\lambda^{2}}\sum_{k=1}^{\infty}\frac{z^{k}}{k^{2}}-g\frac
{S}{\lambda^{2}}\frac{a}{\lambda}\pi g\frac{j+1}{2j+1}\sum\limits_{k=2}%
\sum_{\mu=1}^{k-1}A_{\mu}\frac{z^{k}}{\left(  k-\mu\right)  ^{3/2}}\nonumber\\
&  -g\frac{S}{\lambda^{2}}\left(  \frac{a}{\lambda}\right)  ^{2}\pi
g\frac{j+1}{2j+1}\sum\limits_{k=2}\sum_{\mu=1}^{k-1}\left[  B_{\mu}\frac
{1}{\left(  k-\mu\right)  ^{3/2}}+2A_{\mu}\frac{1}{\left(  k-\mu\right)  ^{2}%
}\right]  z^{k}.\label{i10}%
\end{align}
Then we can calculate $N$:%
\begin{align}
N  &  =g\frac{S}{\lambda^{2}}\sum_{k=1}^{\infty}\frac{z^{k}}{k}-g\frac
{S}{\lambda^{2}}\frac{a}{\lambda}\pi g\frac{j+1}{2j+1}\sum\limits_{k=2}%
\sum_{\mu=1}^{k-1}kA_{\mu}\frac{z^{k}}{\left(  k-\mu\right)  ^{3/2}%
}\nonumber\\
&  -g\frac{S}{\lambda^{2}}\left(  \frac{a}{\lambda}\right)  ^{2}\pi
g\frac{j+1}{2j+1}\sum\limits_{k=2}\sum_{\mu=1}^{k-1}k\left[  B_{\mu}\frac
{1}{\left(  k-\mu\right)  ^{3/2}}+2A_{\mu}\frac{1}{\left(  k-\mu\right)  ^{2}%
}\right]  z^{k}.\label{i11}%
\end{align}
Comparing with (\ref{8}) we obtain the coefficients:%
\begin{align*}
A_{k}  &  =\frac{1}{k}\\
B_{1}  &  =0,\\
B_{k}  &  =-\pi g\frac{j+1}{2j+1}\sum_{\mu=1}^{k-1}\frac{k}{\mu}\frac
{1}{\left(  k-\mu\right)  ^{3/2}}\\
C_{1}  &  =0,\\
C_{k}  &  =\left(  \pi g\frac{j+1}{2j+1}\right)  ^{2}\sum_{\mu=2}^{k-1}%
\sum_{\nu=1}^{\mu-1}\frac{k\mu}{\nu}\frac{1}{\left(  \mu-\nu\right)  ^{3/2}%
}\frac{1}{\left(  k-\mu\right)  ^{3/2}}-2\pi g\frac{j+1}{2j+1}\sum_{\mu
=1}^{k-1}\frac{k}{\mu}\frac{1}{\left(  k-\mu\right)  ^{2}}%
\end{align*}
Substituting these coefficients into Eq. (\ref{i10}), we have%
\begin{align}
\ln\Xi &  =g\frac{S}{\lambda^{2}}\sum_{l=1}^{\infty}\frac{z^{l}}{l^{2}}%
-g\frac{S}{\lambda^{2}}\frac{a}{\lambda}\left(  g\pi\frac{j+1}{2j+1}\right)
\sum_{\mu,l=1}^{\infty}\frac{1}{\mu}\frac{z^{\mu+l}}{l^{3/2}}\nonumber\\
&  +g\frac{S}{\lambda^{2}}\left(  \frac{a}{\lambda}\right)  ^{2}\left(  \pi
g\frac{j+1}{2j+1}\right)  ^{2}\sum_{k,l,\nu=1}^{\infty}\frac{\nu+k}{\nu}%
\frac{1}{k^{3/2}}\frac{z^{\nu+k+l}}{l^{3/2}}-g\frac{S}{\lambda^{2}}\left(
\frac{a}{\lambda}\right)  ^{2}\left(  2\pi g\frac{j+1}{2j+1}\right)  \sum
_{\mu,l=1}^{\infty}\frac{1}{\mu}\frac{z^{\mu+l}}{l^{2}}\nonumber
\end{align}
Therefore, following a similar procedure to that which led to Eq. (\ref{5}),
we obtain the equation of state for the two-dimensional hard-sphere gas, Eq.
(\ref{i06}), again.

\section{Conclusions}

In conclusion, by using a method developed for calculating the boundary effect
we treat a hard-sphere gas as an ideal quantum gas in a multi-core container,
and calculate the grand potentials for hard-sphere gases in two and three
dimensions. The result shows that by only replacing the moving gas molecules
with a fixed boundary, we can obtain the results almost accurate to the second
order, $\left(  a/\lambda\right)  ^{2}$. This is because in this approach we
use the $s$-wave scattering cross-section as the area of the cores. When the
$p$-wave contribution is taken into account, one should consider the effect of
the motion of the cores. Such a simplified model for interacting gases allows
us to calculate the hard-sphere gas in a simple manner, and can be used to
calculate the boundary effects in interacting quantum gases since in this
treatment the hard-sphere interaction is regarded as a boundary effect, which
we will discuss elsewhere. One advantage of this approach is that it can be
used to calculate the hard-sphere gas confined in a container with the
Dirichlet boundary condition since in this paper the method used to calculate
the boundary effect is valid both for periodic and Dirichlet boundary
conditions; while, e.g., in the pseudopotential method the periodic boundary
play an essential role in the calculation since with such a boundary condition
the momentum is conserved \cite{Huang}.\ For seeking more higher-order
solutions, we need to take into account other corrections, e.g., the
correction caused by the movement of the cores, etc. In such a consideration
the grand potential cannot be expanded in powers of $a/\lambda$ as the
higher-order result obtained by using the method of pseudopotentials
\cite{Wu}. This method can be extended to more general cases: by replacing the
diameter of hard-spheres with the scattering length, we can obtain a
generalized result which can be applied to other kinds of interacting gases.
Moreover, this result still can be carried one step further. The expansion
coefficients can be explained as form factors. In the case of hard-sphere
interactions, such form factors are constants. If we introduce a set of
parameter-dependent form factors, the method can be generalized to describe
more complex interacting gases. Such a treatment, which will be discussed in
detail elsewhere, is equivalent to replacing the hard sphere by a soft sphere
which has inner structures. Moreover, this approach can also be directly
applied to quantum gases or liquids in porous media. Porous media can be
regarded as containers with many holes (cores); in this case the particle
number $N$ in Eq. (\ref{1}) should be replaced by the number of such holes
(cores) which are actually real boundaries of the container.

\vskip 0.5cm

We are very indebted to Dr. G. Zeitrauman for his encouragement. This work is
supported in part by the science fund of Tianjin University and Education
Council of Tianjin, P. R. China, under Project No. 20040506. The computation
of this project was performed on the HP-SC45 Sigma-X parallel computer of ITP
and ICTS, CAS.

\section{Appendix\cite{Referee}}

As mentioned above, the equation for $N$ is solvable; in this Appendix we give
the solution for this equation. From Eq. (\ref{1}), the relation
$N=z\partial\ln\Xi/\partial z$ gives a differential equation for $N$,%
\begin{equation}
\frac{\partial N}{\partial z}+\frac{1+\omega\frac{2a}{\lambda}h_{1/2}%
+\omega\frac{2\pi a^{2}}{\lambda^{2}}h_{1}}{\omega z\left(  \frac{2a}{\lambda
}h_{3/2}+\frac{2\pi a^{2}}{\lambda^{2}}h_{2}\right)  }N-\frac{\left(
2j+1\right)  \frac{V}{\lambda^{3}}h_{3/2}}{\omega z\left(  \frac{2a}{\lambda
}h_{3/2}+\frac{2\pi a^{2}}{\lambda^{2}}h_{2}\right)  }=0.\label{A1}%
\end{equation}
The solution of this equation is%
\begin{equation}
N=-e^{\int-p\left(  z\right)  dz}\left[  \int e^{\int p\left(  z\right)
dz}q\left(  z\right)  dz\right]  ,\label{A2}%
\end{equation}
where%
\begin{align*}
p\left(  z\right)   &  =\frac{1+\omega\frac{2a}{\lambda}h_{1/2}+\omega
\frac{2\pi a^{2}}{\lambda^{2}}h_{1}}{\omega z\left(  \frac{2a}{\lambda}%
h_{3/2}+\frac{2\pi a^{2}}{\lambda^{2}}h_{2}\right)  },\\
q\left(  z\right)   &  =-\frac{\left(  2j+1\right)  \frac{V}{\lambda^{3}%
}h_{3/2}}{\omega z\left(  \frac{2a}{\lambda}h_{3/2}+\frac{2\pi a^{2}}%
{\lambda^{2}}h_{2}\right)  }.
\end{align*}
However the form of this solution is complex, and it is difficult to covert
this solution to a simple and more useful form. Of course, the solution given
by Eq. (\ref{A2}) is consistent with the series solution given in the paper.
To check this, we can, for example, expand Eq. (\ref{A2}) and compare it with
the expansion of the result given in the paper. When $z<1$, the expansion of
Eq. (\ref{A2}) and the result of $N$ in the paper both are%
\[
N=\left(  2j+1\right)  \frac{V}{\lambda^{3}}z+\left(  2j+1\right)  \frac
{V}{\lambda^{3}}\left[  \frac{\sqrt{2}}{4}-4\omega\frac{a}{\lambda}-4\pi
\omega\left(  \frac{a}{\lambda}\right)  ^{2}\right]  z^{2}+\cdots.
\]
They are the same. In a word, what we do in the paper is to find a series
solution for the differential equation of $N$.

\end{document}